\documentclass[aps,prl,twocolumn,showpacs,floatfix]{revtex4}
\usepackage{epsfig}
\usepackage{graphicx}
\usepackage{dcolumn}
\usepackage{longtable}
\usepackage{amsthm,amsmath}
\begin{document}

\preprint{\today}

\title{Rigorous limits on the hadronic and semi-leptonic CP-violating coupling constants from the electric dipole moment of $^{199}$Hg}

\author{Yashpal Singh and B. K. Sahoo }
\affiliation{Theoretical Physics Division, Physical Research Laboratory, Navrangpura, Ahmedabad - 380009, India}
\email{yashpal@prl.res.in}       
\email{bijaya@prl.res.in}

\begin{abstract}
Relativistic many-body methods at different levels of approximations are employed to gain insights into the passage of the electron 
correlations from the lower to higher orders in the accurate determination of the electric dipole polarizability ($\alpha_d$) and the
electric dipole moment (EDM) due to the electron-nucleus tensor-pseudotensor (T-PT) and the nuclear Schiff moment (NSM) interactions in
$^{199}$Hg. Moreover, plausible reasons for the differences in the previous atomic calculations are pointed out. Comparison between 
the calculated and experimental results of $\alpha_d$ indicates that our EDM calculations are about 3\% accurate, which in combination 
with the measured value of $^{199}$Hg EDM yield limits on the T-PT coupling constant as $C_T < 2.31 \times 10^{-9}$ and on the NSM as 
$S<1.55 \times 10^{-12}|e|fm^3$. Using these values together with the latest nuclear structure and quantum chromodynamics calculations,
we get limits for the strong CP-violating parameter as $|\bar{\theta}| < 1.1 \times 10^{-9}$ and for the combined up and down quark 
chromo-EDMs as $|\tilde{d}_u - \tilde{d}_d| < 2.9  \times 10^{-26} |e| cm$, which are elevated from their previously known limits. 
\end{abstract}


\maketitle

The existence of the permanent electric dipole moment (EDM) in a non-degenerate system is a signature of parity (P) and time-reversal (T) 
violation (P,T-odd) \cite{landau}. Broken  T symmetry implies CP-violation as a consequence of the CPT theorem \cite{luders}. Violations 
of both the CP and T symmetries have already been observed independently, but only in the hadronic sectors and are well within the predictions of the standard model 
(SM) of particle physics \cite{christenson, alvarez}. The CP-violation in the SM arises through a complex phase parameter $\delta$ of 
the Cabibbo-Kobayashi-Maskawa (CKM) matrix and through the P,T-odd interactions between the quarks and gluons, which is characterized by 
$\bar{\theta}$ parameter \cite{ramsey, barr, pospelov}. However, the observed hadronic CP violation is not sufficient to account for 
the matter-antimatter asymmetry of our Universe \cite{dine, canetti}. Theories like minimal supersymmetric extension of the SM and weak 
scale supersymmetry model (two variants of supersymmetry), multi-Higgs model, the two-loop contributions to SM etc. have numerous new 
sources of CP violation \cite{pospelov, engel, barr1}. These models are also capable of explaining the larger upper limits of
the atomic EDMs. Moreover, the underlying physics describing the atomic EDMs can also support the 
electroweak baryogenesis and can provide information on the light dark matter candidates \cite{kazarian}. 

Although there has not been observation of finite EDM in any system reported so far, but the measurements are continuously improving the 
limits. Currently, the upper limit to the electron EDM ($d_e$) is reported as $d_e < 8.7 \times 10^{-29} |e| cm$ from ThO, an 
open-shell system, EDM measurement which is about ten orders larger than the SM predicted value \cite{baron}. However, EDMs of the 
closed-shell (diamagnetic) atoms are conducive to infer strong CP-odd parameters like $\bar{\theta}$ and EDMs and chromo-EDMs of quarks 
that are predicted by certain supersymmetric and leptoquark models \cite{barr, pospelov, engel, barr1, fukuyama}. Till 
date the best upper limit on the atomic EDM ($d_A$) of a diamagnetic system has been accomplished from $^{199}$Hg as 
$|d_A(^{199}Hg)|< 3.1\times 10^{-29} |e| cm$ \cite{griffith,swallows} with 95\% confidence level (C.L.). 

To determine the limits on the strong CP-odd parameters from the above experimental result, it is imperative to perform the corresponding 
atomic and nuclear calculations reliably. However, not only large discrepancies among the previous atomic calculations are witnessed
in the past \cite{martensson,dzuba02,dzuba09,lathalett}, but the recent nuclear calculations also provide completely different results 
\cite{engel,dekens}. Therefore, all the previously yield limits inferred combining these calculations with the above measurement require
absolute revision. In this Letter, we intent to scrutinize the atomic calculations by employing several many-body methods, including 
those were considered in the previous calculations, and present more reliable results. 

The P,T-odd Lagrangian for the electron-nucleon ($e-n$) interaction is given by \cite{pospelov}
\begin{eqnarray}
\mathcal{L}_{e-n}^{PT} &=& C_T^{e-n} \varepsilon_{\mu \nu \alpha \beta} \bar{\psi}_e \sigma^{\mu \nu} \psi_e  
\bar{\psi}_n \sigma^{\alpha \beta} \psi_n \nonumber \\ && + C_P^{e-n}  \bar{\psi}_e  \psi_e \ \bar{\psi}_n i \gamma_5 \psi_n,
\end{eqnarray}
where $\varepsilon_{\mu \nu \alpha \beta}$ is the Levi-Civita symbol, $\sigma_{\mu \nu} = \frac{i}{2}[\gamma_\mu, \gamma_\nu]$ with $\gamma$s as 
the usual Dirac matrices, $C_T^{e-n}$ and $C_P^{e-n}$ are the tensor-pseudotensor (T-PT) and scalar-pseudoscalar (S-PS)
$e-n$ interaction coupling constants, respectively, and $\psi_{n/e}$ representing the Dirac wave functions for the corresponding 
particle $n/e$. Assuming that the T-PT and S-PS interactions act independently, we can consider them individually in the atomic 
calculations for which the respective electron-nucleus ($e-N$) interaction Hamiltonians in an atomic system 
are given as \cite{dzuba09}
\begin{eqnarray}
 && H_{e-N}^{TPT} = i \sqrt{2} G_F C_T \sum_e \mbox{\boldmath $\sigma_N \cdot \gamma$} \rho_N(r) \\
\text{and} && \nonumber \\
&& H_{e-N}^{SPS} = - \frac{G_F}{\sqrt{2}m_n c} C_P \sum_e \gamma_0 \mbox{\boldmath $\sigma_N \cdot \nabla$} \rho_N(r), 
\end{eqnarray}
where $G_F$ is the Fermi constant, $C_T$ and $C_P$ are the $e-N$ T-PT and S-PS coupling constants, {\boldmath$\sigma_N$}$=\langle 
\sigma_N \rangle \frac{{\bf I}}{I}$ is the Pauli spinor of the nucleus with spin $I$, $\rho_N(r)$ is the nuclear density, $m_n$ is the 
mass of the odd nucleon and $c$ is the speed of light. $C_P$ can be estimated within reasonable accuracy using an empirical 
relation \cite{dzuba09}
\begin{eqnarray}
C_P \approx 3.8 \times 10^3 \times \frac{A^{1/3}}{Z} C_T
\label{eqcp}
\end{eqnarray}
with the atomic number $Z$ and mass $A$. Hence, we only consider the T-PT $e-N$ interaction in our calculation.

The Lagrangian for the other dominant P,T-odd pion-nucleon-nucleon ($\pi-n-n$) interaction in the diamagnetic atoms is given by \cite{pospelov}
\begin{eqnarray}
\mathcal{L}^{\pi n n}_{e-n} &=& \bar{g}_0 \bar{\psi}_n \tau^i \psi_n \pi^i + \bar{g}_1 
\bar{\psi}_n \psi_n \pi^0 \nonumber \\ && + \bar{g}_2 \big ( \bar{\psi}_n \tau^i \psi_n \pi^i - 
3 \bar{\psi}_n \tau^3 \psi_n \pi^0 \big )
\end{eqnarray}
where $\bar{g}_{i=0,1,2}$s are the CP-odd $\pi-n-n$ couplings and $\tau^{i=1,2,3}$s are the isospin components. The corresponding $e-N$ interaction 
Hamiltonian is given by \cite{dzuba09}
 \begin{eqnarray}
  H_{e-N}^{NSM}= \frac{3{\bf S.r}}{B_4} \rho_N(r),
 \end{eqnarray}
where ${\bf S}=S \frac{{\bf I}}{I}$ is the NSM and $B_4=\int_0^{\infty} dr r^4 \rho_N(r)$. 

When the above interactions are taken along with the Dirac-Coulomb (DC) atomic Hamiltonian ($H_a$), the ground state wave function $|\Psi_0 \rangle$ of 
the atom becomes admixture of opposite parities. The $d_A$ of this state, the expectation value of the electric dipole operator $D$, is evaluated 
in the first order approximation \cite{martensson,dzuba02,dzuba09,lathalett, yashpal-rapid} as
\begin{eqnarray}
 d_A &=& 2 \frac{\langle \Psi_0^{(0)}|D|\Psi_0^{(1)} \rangle}{\langle \Psi_0^{(0)}|\Psi_0^{(0)} \rangle},
 \label{eqed}
\end{eqnarray}
where $| \Psi_0^{(0)} \rangle$ and $|\Psi_0^{(1)} \rangle$ are the wave functions of $H_a$ and its first order correction due to 
the P,T-odd interaction Hamiltonian, respectively. Since the rank and parity of $D$ are same as of the considered weak interaction 
Hamiltonians, therefore insights into the accuracy of $d_A$ can be provided by calculating $\alpha_d$, for which $|\Psi_0^{(1)} \rangle$ 
need to be evaluated by perturbing $|\Psi_0^{(0)} \rangle$ by $D$ in Eq. (\ref{eqed}). We consider the DF wave function, $|\Phi_0\rangle$, as the reference wave 
function to start with and then the neglected electron correlation effects are included using the second (MBPT(2)) and third (MBPT(3)) 
order many-body perturbation theory, random phase approximation (RPA) and coupled-cluster (CC) theory using the normal ordered $H_a$
with respect to $|\Phi_0 \rangle$. We have explained implementations of these methods in detail elsewhere for the calculations of 
$\alpha_d$ and $d_A$ \cite{yashpal-polz,yashpal-arxiv,yashpal-rapid}. In our relativistic CC calculations, we have considered only the 
single and double excitations, denoted by the subscripts 1 and 2 respectively, retaining the linear terms (LCCSD method) as well as 
accounting for all the linear and non-linear terms (CCSD method). As will be demonstrated later, we find there are large differences 
between our results with another calculations of the above quantities for $^{199}$Hg \cite{lathalett, siddhartha} using an analogous CC 
method (called as PRCC) that includes contributions from the normalization factor $\langle \Psi_0^{(0)}|\Psi_0^{(0)} \rangle$. It can
be shown that the numerator of an expectation value determining expression of an normal ordered operator in the ground state will have two disjoint
closed parts with and without the operator in the CC method and the part without the operator will cancel out with the normalization 
factor \cite{cizek, bartlett}. Here, we demonstrate that there is also same cancellation occurs with the normalization factor when the matrix 
element of an normal ordered operator between the unperturbed and the first-order perturbed wave functions, as in Eq. (\ref{eqed}), 
is evaluated.

In a CC approach, we express the atomic wave function due to $H_a$ and its first order correction due to the P,T-odd interaction 
Hamiltonians or $D$ as \cite{lathalett, yashpal-polz,yashpal-arxiv,yashpal-rapid, siddhartha}
\begin{eqnarray}
 | \Psi_0^{(0)} \rangle = e^{T^{(0)}} |\Phi_0 \rangle
 \ \ \ \ \text{and} \ \ \ \ 
 | \Psi_0^{(1)} \rangle = e^{T^{(0)}} T^{(1)} |\Phi_0 \rangle,
\label{eq33}
\end{eqnarray}
respectively, where the operators $T^{(0)}$ creates even and $T^{(1)}$ creates odd parity excitations from $|\Phi_0 \rangle$. 
Then, $d_A$ or $\alpha_d$ (both are denoted by $X$) is evaluated by  
\begin{eqnarray}
 X &=& 2 \frac{\langle\Phi_0 | e^{T^{\dagger (0)}} D e^{T^{(0)}} T^{(1)} | \Phi_0 \rangle }
                  {\langle\Phi_0 | e^{T^{\dagger (0)}} e^{T^{(0)}} | \Phi_0 \rangle }.
\end{eqnarray}
Since all the operators in the above expression are in normal order form and $e^{T^{\dagger (0)}} D e^{T^{(0)}}$ is a non-truncative series,
we can express $e^{T^{\dagger (0)}} D e^{T^{(0)}}= (e^{T^{\dagger (0)}} e^{T^{(0)}})_{cl} (e^{T^{\dagger (0)}} D e^{T^{(0)}})_{cc}$ with 
the subscript $cl$ means the terms are closed and the subscript $cc$ means the terms are connected and closed \cite{bartlett, pal}. 
Following similar approaches, we can show that
\begin{eqnarray}  
X &=& 2 \frac{\langle\Phi_0 | (e^{T^{\dagger (0)}} e^{T^{(0)}})_{cl} (e^{T^{\dagger (0)}} D e^{T^{(0)}} T^{(1)})_{cc} | \Phi_0 \rangle }
                  {\langle\Phi_0 | (e^{T^{\dagger (0)}} e^{T^{(0)}})_{cl} | \Phi_0 \rangle } \nonumber \\
  &=& 2 \frac{\langle\Phi_0 | (e^{T^{\dagger (0)}} e^{T^{(0)}})_{cl} |\Phi_0 \rangle \langle \Phi_0 | (e^{T^{\dagger (0)}} D e^{T^{(0)}} T^{(1)})_{cc} | \Phi_0 \rangle }
                  {\langle\Phi_0 | (e^{T^{\dagger (0)}} e^{T^{(0)}})_{cl} | \Phi_0 \rangle } \nonumber \\
  &=& 2 \langle\Phi_0 |(\overline{D}^{(0)} T^{(1)})_{cc}|\Phi_0 \rangle,
\label{eq38}
\end{eqnarray}
with $\overline{D}^{(0)} = e^{T^{\dagger{(0)}}}De^{T^{(0)}}$, which is also a non-truncative series. Note that its 
$(e^{T^{\dagger (0)}} e^{T^{(0)}}T^{(1)})_{cl}$ part will vanish owing to odd-parity nature of $T^{(1)}$. In the LCCSD method, we get 
$\overline{D}^{(0)} = D + DT^{(0)} + T^{\dagger{(0)}} D + T^{\dagger{(0)}} D T^{(0)}$. To account contributions from 
$\overline{D}^{(0)}$ in the CCSD method, we first evaluate terms from $\overline{D}^{(0)}$ that are very unique in the sense that they 
will not be repeated when these terms are further contracted with another $T^{(0)}$ or $T^{\dagger (0)}$ operator. Contributions 
from the higher non-linear terms are accounted by contracting the above dressed effective operators with another $T^{(0)}$ 
and $T^{\dagger (0)}$ operators till the self-consistent results are achieved. We present these contributions with $k$ numbers of 
$T^{(0)}$ and or $T^{\dagger (0)}$ as CCSD$^{(k)}$ to demonstrate convergence in the results with the series.

\begin{table}[t]
\caption{Results of $\alpha$, $d_A^{T}$, $d_A^{S}$ in units $e a_0^3$, ($10^{-20}C_T \langle \sigma\rangle |e|cm$) and ($10^{-17}[S/|e|fm^3]|e|cm$) respectively are
presented using different many body methods for the ground state of $^{199}$Hg and compared them with others calculations. Refs.
$^a$\cite{dzuba02} ($^\dagger$ also contains RPA contributions), $^b$\cite{martensson}, $^c$\cite{dzuba09}, $^d$\cite{pershina}, 
$^e$\cite{lathalett}, $^f$\cite{siddhartha} ,  $^g$\cite{goebel}.}
\begin{ruledtabular}
\begin{tabular}{lcccccc}
\textrm{Method}& \multicolumn{3}{c}{\textrm{This Work}}   &\multicolumn{3}{c}{\textrm{Others}} \\
 \cline{2-4}  \cline{5-7}\\
               & $\alpha_d$&  $d_A^{T}$& $d_A^{S}$ &$\alpha_d$ &$d_A^{T}$ & $d_A^{S}$\\
              \hline\\
DF     & 40.95 &$-$2.39  &$-$1.20   &40.91$^a$&$-$2.0$^b$  & $-$1.19$^a$ \\
       &       &         &          &44.90$^d$& $-$2.4$^c$&$-$1.2$^c$   \\
MBPT(2)& 34.18  &$-$4.48  &$-$2.30   &   &                             &\\    
MBPT(3)& 22.98  &$-$3.33  &$-$1.72   &   &                             &\\       
RPA    & 44.98  &$-$5.89  &$-$2.94   &44.92$^a$   &$-$6.0$^b$& $-$2.8$^a$\\
       &       &         &          &   &$-$5.9$^c$    & $-$3.0$^c$       \\
CI+MBPT&       &         &          &$^\dagger$32.99$^a$   &$-$5.1$^c$ & $-$2.6$^c$ \\
PRCC   &       &         &          &33.294$^e$  &    $-$4.3$^e$& $-^*$5.07$^e$ \\  
       &       &         &          &33.59$^f$&&\\
LCCSD  &33.91  &$-$4.52  &$-$2.24   &   &                             &\\
CCSD$^{(2)}$ &33.76  &$-$3.82  &$-$2.00   &   &                             &\\
CCSD$^{(4)}$ &35.13  &$-$4.14  &$-$2.05   &   &                             &\\
CCSD$^{(\infty)}$ &34.98  &$-$4.02  &$-$2.00   & 35.31$^d$ &  & \\              
Expt.    &33.91(34)$^g$&      & \\
\end{tabular}
\end{ruledtabular}  
\label{tab1}
\footnotetext{* We learned that the correct value is $-2.46$ \cite{dilip}.}
\end{table}  
In Table \ref{tab1}, we summarize the results of $\alpha_d$ and $d_A$ from the calculations obtained using different levels of 
approximations in the many-body methods and from the measurement. It is worthwhile to note that the time dependent Hartree-Fock 
(TDHF) method of Refs. \cite{dzuba02,dzuba09} and coupled Hartree-Fock of method Ref. \cite{martensson} are basically same as our 
RPA and hence the results from all these methods are reported in the same row. Similarly, the results reported by the PRCC methods 
\cite{lathalett, siddhartha} are similar to our CC methods, but they differ in the procedure of determining amplitudes of the CC 
operators and evaluating Eq. (\ref{eq38}) in the final property calculations \cite{dilip}. 

\begin{table}[t]
\caption{Contributions to the $\alpha$ in $e a_0^3$, $d_A^T$ and $d_A^{S}$ values from various CCSD terms (hermitian conjugate 
terms are included). Here $norm$ represents difference between the contributions after and before normalizing the wave function with 
normalization factor 1.171 and $NA$ stands for not applicable.}
\begin{ruledtabular}
\begin{tabular}{lcccc}
CC &  \multicolumn{3}{c}{ This work}  & Ref. \cite{siddhartha}\\
\cline{2-4}& \\
term & $\alpha_d $& $d_A^T$ & $d_A^{S}$ & $\alpha_d$\\
\hline        \\
$DT_1^{(1)}$                &  39.77   &$-$5.00  & $-$2.44 &41.927\\
$T_2^{(0)\dagger}DT_1^{(1)}$& $-$5.73  &   1.36  &    0.62 &$-$2.724\\
$T_2^{(0)\dagger}DT_2^{(1)}$&   1.55   &$-$0.11  & $-$0.06 &1.504\\
$T_1^{(0)\dagger}DT_1^{(1)}$& $-$1.71  &   0.02  &    0.02 &$-$1.583\\
$T_1^{(0)\dagger}DT_2^{(1)}$& $-$0.12  &$-$0.08  &    0.04 &0.091\\
$Extra$                     &   1.22   &$-$0.21  & $-$0.18 &0.119\\
$norm$            &   $NA$    &  $NA$     &   $NA$   &$-$5.74\\
\hline 
& & \\
Total                     &   34.98  &$-$4.02  & $-$2.00 &33.59\\
\end{tabular} 
\end{ruledtabular}
\label{tab2}
\end{table}
 Categorically, two different approaches are adopted to calculate $\alpha_d$ among which Pershina {\it et al.} \cite{pershina} evaluate
the second derivative of the ground state energy with respect to an arbitrary electric field and the other calculations 
\cite{dzuba02,lathalett, siddhartha} determine the expectation value of $D$ in the ground state which has a mixed parity wave function.
Since the DF method gives upper bound to the exact energy, Pershina {\it et al.} get a large DF value and their CCSD method brings
down the result towards the experimental value. Furthermore, they show that inclusion of the important triple excitations in their CCSD 
calculations gives rise 34.15 $ea_0^3$ which is almost in agreement with the experimental result. To compare the results between
our calculations with that are reported in \cite{lathalett, siddhartha}, we present contributions from different CC terms from the
CCSD method for all the evaluated quantities in Table \ref{tab2}. We, however, learned that there are large differences in the results 
of Ref. \cite{lathalett} and Ref. \cite{siddhartha} with the DC Hamiltonian than the published values due to 
wrong phase factors in the adjoint cluster operators \cite{dilip}, although both the references report very close values for $\alpha_d$. In Ref. \cite{siddhartha}, 
$\alpha_d$ result has been improved by including contributions from the Breit interaction and important triple excitations \cite{dilip}.
Nevertheless, comparison between these calculations in Table \ref{tab2} shows large differences in the contributions among individual CC terms 
and from the normalization factor of the wave function, which cancels out naturally in our method. Also, we notice significant 
differences in the contributions from the non-linear terms arising in Eq. (\ref{eq38}) in both the CC approaches, which are given as 
$Extra$ in Table \ref{tab2}, even though both the works consider only the connected diagrams in the calculations. On the other hand, our DF 
and RPA $\alpha_d$ results agree quite well with that are reported in Refs. \cite{dzuba02, dzuba09}. From the differences between our 
final CCSD and experimental results \cite{goebel} of $\alpha_d$, we estimate that our reported EDM results are about 3\% accurate. 

The main aim of this Letter is to find out more reliable EDM results for $^{199}$Hg, where two previous calculations using the PRCC method 
and a hybrid approach of configuration interaction with finite-order many-body perturbation theory (CI+MBPT) differ substantially
\cite{lathalett, dzuba09} as seen in Table \ref{tab1}. In the CI+MBPT method, the initial wave functions are determined using the $V^{N-2}$
potential with $N$ as the total number of electrons and the electron correlation effects are accounted for by dividing the electrons 
into valence and core electrons. In contrast, both the PRCC method and our calculations are carried out using the $V^N$ potential and 
correlations among all the electrons are treated on equal footing. We find, like the $\alpha_d$ results, our DF and RPA results for 
$d_A$ due to the T-PT ($d_A^T$) and NSM ($d_A^S$) interactions match perfectly with Refs. \cite{dzuba02,dzuba09} and also with 
another old calculation \cite{martensson}. The large differences between the results of Refs. \cite{lathalett} and \cite{dzuba09} may be 
attributed to the phase factor problem \cite{dilip}, as discussed above in the context of $\alpha_d$ result. In an earlier calculation,
Dzuba {\it et al.} had also reported $\alpha_d$ as 32.99 $ea_0^3$ using the CI+MBPT method along with some corrections from RPA
\cite{dzuba02}, which also differs by about 3\% from its experimental value but in the lower side. We recommend our final EDM results
from the CCSD method, which are obtained from Eq. (\ref{eq38}) using an iterative procedure, are the most accurate and reliable 
on the physical grounds. The reason is that even our LCCSD and CCSD$^{(2)}$ results for $\alpha_d$ are very close with its experimental 
result, but the CCSD method includes more physical effects from both the RPA and non-RPA correlation effects through the $T_1^{(0)} 
T_2^{(0)}$, $\frac{1}{2} T_2^{(0)} T_2^{(0)}$, $\cdots$ non-linear terms that corresponds to the higher level excitations. Similarly
our MBPT(2) results, which are the lowest order RPA, seem to be closer to the experimental result and with some of the all order 
calculations, but they pose serious doubt over their accuracy owing to the fact that higher correlation corrections are non-negligible.
The importance of the non-RPA effects can be found from the differences between the MBPT(3) (where the non-RPA contributions start 
entering in the perturbation theory) and MBPT(2) results at the lowest order level and from the differences between the CCSD and RPA results at the all order 
level. They were almost canceled out with the higher order RPA contributions in $^{129}$Xe \cite{yashpal-rapid} and $^{229}$Rn 
\cite{sahoo}, but here they are found to be very crucial in order to get accurate results.

Combining our final CCSD results with the measurement \cite{griffith, swallows}, we get the most accurate bounds as
\begin{eqnarray}
 S \ \textless \ 1.55 \times 10^{-12} |e|fm^3 \ \ \ \text{and} \ \ \ C_T \ \textless \ 2.31 \times 10^{-9} .
\label{eq11}
\end{eqnarray}
Using Eq. (\ref{eqcp}), it yields $C_P<6.4\times 10^{-7}$ and from the relation $S=s_pd_p+s_nd_n$ for the Landau-Migdal parameters 
$s_p=0.20\pm0.02$ fm$^2$ and $s_n=1.895\pm0.035$ fm$^2$ \cite{dmitriev}, we get the limits on the neutron ($d_n$) and proton ($d_p$) EDMs as
\begin{eqnarray}
 d_n \ \textless \ 8.18 \times 10^{-26} |e|cm  \  \text{and} \ 
 d_p \ \textless \ 7.75 \times 10^{-25} |e|cm . \ \
\end{eqnarray}
Although our extracted limit on $d_n$ is not better than the limit obtained from the direct measurement \cite{baker}, but 
the limit on $d_p$ estimated to be better than the previous value \cite{griffith}. 

In a recent review, Engel {\it et al} discuss about large uncertainties in the nuclear calculations and discordance among all these 
results both in signs and magnitudes for $^{199}$Hg and recommend the best value for $S$ as \cite{engel}
\begin{eqnarray}
 {\boldmath S}&=& 13.5[ 0.01 \bar{g}_0 + (\pm 0.02) \bar{g}_1 + 0.02 \bar{g}_2]\mbox{ $|e|$ fm}^3.
\end{eqnarray}
Combining this with our limit on $S$, we infer bounds as $|\bar{g}_0|\textless 1.2\times 10^{-11} $ and $|\bar{g}_1|\textless 5.74\times 
10^{-12}$. Further using the relations $\bar{g}_0=(-0.018\pm 0.007) \bar{\theta}$ \cite{dekens} and $\bar{g}_1=2\times 10^{-12}
(\tilde{d}_u-\tilde{d}_d)$ \cite{pospelov1}, we extract the upper limit on the combined up and down quarks chromo-EDMs as $|\tilde{d}_u - 
\tilde{d}_d| < 2.9 \times 10^{-26} |e| cm$ and the limit on the strong CP-odd parameter as $|\bar{\theta}|\textless 1.1\times 10^{-9}$. 
In fact, it is also possible to put more stringent limits on the above quantities from our given limit on $S$ provided the uncertainties in 
the nuclear calculations are reduced further.

The authors acknowledge Professor B. P. Das, Dr N. Mahajan and Dr J. de Vries for many useful discussions. 
The computations were carried out using 3TFLOP HPC cluster of PRL, Ahmedabad.

\end{document}